\documentclass{egpubl}
\usepackage{eurovis2025}

\SpecialIssuePaper         

\CGFccby

\usepackage[T1]{fontenc}
\usepackage{dfadobe}  
\PassOptionsToPackage{svgnames}{xcolor}
\usepackage{tikz} 
\usetikzlibrary{calc,intersections}
\usepackage{adjustbox}
\usepackage{tcolorbox} 

\usepackage{booktabs} 
\usepackage{cite}  
\BibtexOrBiblatex
\electronicVersion
\PrintedOrElectronic

\usepackage{egweblnk}

\title[EG \LaTeX\ Author Guidelines]%
      {Benchmarking Visual Language Models on Standardized Visualization Literacy Tests
      }

\author[S. Pandey \& A. Ottley]
{\parbox{\textwidth}{\centering Saugat Pandey\orcid{0000-0002-7429-6575}
        and Alvitta Ottley\orcid{0000-0002-9485-276X} 
        }
        \\
{\parbox{\textwidth}{\centering Washington University in St.\ Louis, St.\ Louis, MO, USA
       }
}
}

\newcommand{\Claude}[0]{ \textsc{\textbf{claude}}}
\newcommand{\Gemini}[0]{ \textsc{\textbf{gemini}}}
\newcommand{\GPT}[0]{ \textsc{\textbf{gpt}}}
\newcommand{\Llama}[0]{ \textsc{\textbf{llama}}}
\newcommand{\CALVI}[0]{\textsc{\textbf{calvi}}}
\newcommand{\VLAT}[0]{\textsc{\textbf{vlat}}}

\begin{document}


\maketitle
\begin{abstract}
The increasing integration of Visual Language Models (VLMs) into visualization systems demands a comprehensive understanding of their visual interpretation capabilities and constraints. While existing research has examined individual models, systematic comparisons of VLMs' visualization literacy remain unexplored. We bridge this gap through a rigorous, first-of-its-kind evaluation of four leading VLMs (GPT-4, Claude, Gemini, and Llama) using standardized assessments: the Visualization Literacy Assessment Test (VLAT) and Critical Thinking Assessment for Literacy in Visualizations (CALVI). Our methodology uniquely combines randomized trials with structured prompting techniques to control for order effects and response variability - a critical consideration overlooked in many VLM evaluations. Our analysis reveals that while specific models demonstrate competence in basic chart interpretation (Claude achieving 67.9\% accuracy on VLAT), all models exhibit substantial difficulties in identifying misleading visualization elements (maximum 30.0\% accuracy on CALVI). We uncover distinct performance patterns: strong capabilities in interpreting conventional charts like line charts (76-96\% accuracy) and detecting hierarchical structures (80-100\% accuracy), but consistent difficulties with data-dense visualizations involving multiple encodings (bubble charts: 18.6-61.4\%) and anomaly detection (25-30\% accuracy). Significantly, we observe distinct uncertainty management behavior across models, with Gemini displaying heightened caution (22.5\% question omission) compared to others (7-8\%). These findings provide crucial insights for the visualization community by establishing reliable VLM evaluation benchmarks, identifying areas where current models fall short, and highlighting the need for targeted improvements in VLM architectures for visualization tasks. To promote reproducibility, encourage further research, and facilitate benchmarking of future VLMs, our complete evaluation framework, including code, prompts, and analysis scripts, is available at \url{https://github.com/washuvis/VisLit-VLM-Eval}.


\end{abstract}  
\section{Introduction}
\label{sec:intro}
Large Language Models (LLMs) and Visual Language Models (VLMs) are emerging tools in data analysis and visualization, attracting considerable attention for their potential to address challenges faced by researchers and practitioners~\cite{ma2023insightpilot, hong2024data}.  
One of the most promising aspects of LLMs and VLMs is their ability to assist individuals with low vision, making visual data more accessible and understandable~\cite{liew2022using,strobel2024hey}. 
They can also help mitigate information overload~\cite{tian2024chartgpt,wang2024aligned} or enable individuals who lack formal training to generate visualizations and engage with data through natural language, asking questions and receiving answers in a conversational manner\cite{choe2024enhancing,sultanum2023datatales,strobel2024hey}. This democratization of data interaction can have a significant impact on how users interact with and interpret information.

Yet, while VLMs offer exciting prospects, there remain substantial questions about their capabilities, limitations, and reliability in visualization tasks, particularly when compared to human performance. Recent research activities in the visualization community are driven by the following questions: \textit{To what extent can VLMs effectively interpret and reason about visual information?} \textit{How do their visualization literacy and perception abilities compare to those of humans?} and \textit{Can VLMs be trusted for use in complex data-driven environments?} These questions are essential because understanding VLMs' strengths and weaknesses could guide their application and help establish whether they are suitable for real-world use or require further development to meet needs.

To address these questions, recent work in the visualization community has examined VLM capabilities, including efforts to replicate foundational perceptual studies and explore applications such as natural language interaction with data, automatic captioning, visualization generation for research and educational purposes, and dataset creation for visualization tasks \cite{guo2024understanding, vazquez2024llms, lo2024good, chen2024viseval, wang2024aligned, kavaz2023chatbot, maddigan2023chat2vis}. These early studies offer valuable insights, but they also reveal significant limitations. First, some studies focus on single VLMs, overlooking the diversity in model architecture and training approaches. Second, while several methods have been proposed to assess people's visualization literacy \cite{lee2016vlat, ge2023calvi, borner2016investigating, boy2014principled, pandey2023mini, cui2023adaptive}, there is limited consensus on standardized benchmarks for VLM evaluation. Furthermore, VLMs exhibit unique challenges, such as a tendency to hallucinate or produce unreliable outputs that are highly sensitive to prompt phrasing and order effects, calling into question their robustness in visualization tasks \cite{ji2023survey, liu2023trustworthy}.

To expand upon this crucial line of research, we present a comprehensive and detailed comparative analysis of four of the most recognized VLMs --- GPT-4, Gemini, Claude, and Llama --- using two standardized visualization literacy assessments: Visualization Literacy Assessment Test (VLAT) \cite{lee2016vlat} and Critical Thinking Assessment for Literacy in Visualizations (CALVI) \cite{ge2023calvi}. 

These assessments are meticulously designed to evaluate essential skills in reading, interpreting, and reasoning with various forms of visual representations, offering a nuanced understanding of each VLM's competencies. Additionally, we compare the performance of these VLMs against established human benchmarks to scrutinize how well these models respond to varying task types, adapt to different visualization styles, and interpret potentially misleading design elements. To ensure the robustness of our findings, we average the results over ten randomized trials to control for any order effects or prompt sensitivities. Through this detailed investigation, we aim to provide a clearer picture of VLMs' potential and limitations in advancing the data visualization field.
We make the following contributions toward understanding the strengths and limitations of VLMs for visual data interpretation:

\begin{itemize}
\item We show that VLMs can accurately analyze a range of visual data formats. 
However, our findings also highlight that the effectiveness of these models varies significantly, influenced by the specific task, visualization type, and the models themselves. Among the models assessed, Claude stood out, demonstrating superior performance.
\item Our comparative analysis with human performance reveals both promising capabilities and concerning limitations. VLMs approach or exceed human-level performance in specific tasks like trend identification (75-80\% accuracy compared to 70\% human average) and hierarchical structure detection (80-100\% accuracy compared to 90\% human average).
\item However, while powerful, they require careful consideration before deployment in complex data-driven environments. Their strong performance in basic chart interpretation but poor reliability in detecting visualization deception (maximum 30.0\% accuracy on CALVI compared to 39\% human average) suggests they are better suited as assistive tools rather than autonomous systems. 
\item We provide a reproducible evaluation framework with randomized trials and structured prompting techniques to help assess future VLM capabilities.
\end{itemize}

\section{Related Works}
Large Language Models (LLMs) and Visual Language Models (VLMs) have transformed artificial intelligence by bridging textual and visual understanding. While LLMs excel in text processing, VLMs expand these capabilities through advanced transformer architectures \cite{bordes2024introduction}, enabling sophisticated visual question answering \cite{guo2023images, hu2024bliva}, image captioning \cite{liu2024visual, chan2023clair}, and multimodal capabilities \cite{lu2022learn, openai2024gpt4technicalreport}. 
These capabilities evolved from Vaswani et al.'s \cite{vaswani2017attention} transformer architecture, progressing through GPT-1 \cite{radford2018improving} to GPT-3 \cite{brown2020language}. The field advanced further with frameworks like VisualBERT \cite{li2019visualbert} and VilBERT \cite{lu2019vilbert}, while models like CLIP \cite{radford2021learning} and Flamingo \cite{alayrac2022flamingo} demonstrated remarkable zero-shot capabilities in visual tasks.

However, these systems face limitations in visual reasoning and multi-modal learning. Key challenges include hallucinations --- where models generate convincing but factually incorrect outputs --- and sensitivity to prompt variations \cite{ji2023survey}. Training data bias presents another concern, as web-scale datasets can perpetuate biases and misinformation \cite{bender2021dangers}. Current research focuses on improving model transparency and reliability through adversarial training and debiasing algorithms \cite{wang2022fairness, shah2024comprehensive}.
\subsection{VLMs in Visualization Research}
Despite the limitations mentioned earlier, research at the intersection of VLMs and visualization has expanded rapidly in recent years, encompassing both the use of visualization techniques to understand and improve VLMs and the application of VLMs to advance visualization systems and tools. Our work contributes to this second research direction, where VLMs enable novel visualization capabilities and applications. By understanding these strengths and limitations, researchers and practitioners can better position LLMs and VLMs for real-world applications while addressing their inherent challenges. 

Recent studies have shown that VLMs have a wide range of applications in visualization, from automated generation of visualization code and charts to sophisticated natural language interfaces for visual analytics systems \cite{gorniak2024vizability, chen2024viseval, tian2024chartgpt}. These applications promise to enhance data literacy by making complex visualizations more accessible through natural language interaction and automated guidance \cite{choe2024enhancing}. However, this rapid adoption has paralleled an increasing focus on rigorous evaluation. Understanding their capabilities, limitations, and reliability becomes crucial as these models become more integrated into visualization systems. The scope and quick expansion of VLM applications in visualization underscore the importance of thorough, systematic evaluation approaches to ensure effective and responsible deployment in real-world scenarios.

\subsection{Visualization Interpretation and Understanding in VLM Research}
Most relevant to the current work, scholars have sought to explore the capabilities of VLMs for tasks related to visualization. There has been a series of recent studies addressing very similar research questions\cite{bendeck2024empirical, guo2024understanding, vazquez2024llms, lo2024good, chen2024viseval}. Most relevant to this work, Bendeck et al. \cite{bendeck2024empirical} presented an empirical investigation of GPT-4's visualization literacy tasks using VLAT, examining performance across 8 different types of tasks across 12 visualization types. They demonstrated that while the model excels at trend identification and design best practices, it struggles with precise value retrieval and color discrimination, with GPT-4's overall performance in the 16th percentile compared to humans. Similarly, Guo et al. \cite{guo2024understanding} investigated VLMs' perceptual capabilities through a series of graphical perception tasks, finding that VLMs can successfully replicate human perceptual judgments, particularly in tasks involving relative comparisons and trend analysis. While this might seem to contrast with Bendeck et al.'s findings, the difference lies in the type of tasks being evaluated - Guo et al. \cite{guo2024understanding} focused specifically on perceptual judgment tasks. At the same time, VLAT encompasses a broader range of visualization interpretation skills.

Other work has sought to examine how well VLMs can identify misleading aspects of charts in addition to the basic ability to read and understand charts. Islam et al. \cite{islam2024large} conducted a comprehensive evaluation of LVLMs across five major chart reasoning tasks, including chart question answering, summarization, and fact-checking. Their findings highlighted that while LVLMs demonstrate strong natural language generation capabilities, they also exhibit common issues such as hallucinations and data bias. In particular, Lo et al. \cite{lo2024good} investigated VLMs' ability to detect misleading visualizations through an exploratory evaluation approach using a dataset of 21 distinct chart issues. While their work showed promising potential for these models in supporting critical thinking during data interpretation, their open-ended methodology of explicitly asking about misleaders highlighted the need for more structured evaluation frameworks. In many real-world scenarios, identifying misleaders is not as straightforward. We take a different approach using CAVI~\cite{ge2023calvi}, which uses targeted questions addressing specific aspects of visual data with clearly defined response options, including an option to indicate when answers are impossible. This method facilitates a more natural identification of misleaders and accommodates the complexities inherent in interpreting visual information.

Our approach overcomes key limitations of earlier studies by controlling for order effects and prompt sensitivities through randomization. This establishes a reproducible framework for future evaluations as VLM technology continues to evolve. Our contribution is particularly timely, given the rapid advancement of VLMs and the growing need for reliable benchmarks in visualization research. 

\subsection{Assessment Frameworks for Visualization Literacy}
The visualization community has developed several approaches for measuring and understanding visualization literacy. Early work by Börner et al. \cite{borner2016investigating} pioneered the examination of public visualization literacy through familiarity-based questions about various data visualizations, revealing important insights about non-expert comprehension patterns. Boy et al. \cite{boy2014principled} established foundational methods using item response theory (IRT), though their work focused primarily on basic chart types like line graphs, bar charts, and scatterplots. 

Building on these foundations, Lee et al. \cite{lee2016vlat} developed the Visualization Literacy Assessment Test (VLAT), offering comprehensive evaluation across 12 visualization types through 53 multiple-choice items. Mini-VLAT \cite{pandey2023mini} later emerged as a more practical assessment tool while maintaining strong psychometric properties. Ge et al. \cite{ge2023calvi} advanced the field by developing CALVI specifically for assessing critical thinking about misleading visualizations, introducing systematic evaluation of ``misleaders'' through 45 validated items. Recent work by Cui et al. \cite{cui2023adaptive} has explored adaptive testing approaches, demonstrating comparable reliability with reduced question sets.

While these assessments have proven effective for evaluating human visualization literacy, there remains a critical gap in their application to VLMs. Despite the increasing use of VLMs in visualization tasks, there is a lack of standardized benchmarking and robust frameworks to assess these technologies across different visualization types and tasks. Our work addresses this gap by leveraging VLAT and CALVI as established benchmarks, extending their application to evaluate VLMs' visualization literacy.
\section{Methodology}
Our methodology addresses three fundamental questions raised in \autoref{sec:intro} by investigating the extent of VLMs' visual interpretation abilities, their comparative performance against human benchmarks, and their reliability in real-world visualization tasks. This comprehensive evaluation framework employs standardized assessments to examine fundamental visualization comprehensive (ability to read and interpret) and critical thinking capabilities (ability to detect misleading visualizations) across different VLMs.

\subsection{Model Selection and Configuration}
Our study focuses on the state-of-the-art VLMs representing different visual-language understanding approaches. We selected these models based on several practical considerations: First, they offer stable, well-documented APIs that support consistent interaction patterns, which are crucial for reproducible research. Second, they can be deployed through cloud-based interfaces, eliminating the need for specialized hardware and making our evaluation framework accessible to a broader research community. Third, these models are widely used in practical scenarios, from data exploration to educational settings, making their evaluation particularly relevant for real-world applications. We excluded models such as Microsoft's CoPilot since it utilizes GPT-4 as its underlying model. Our final selection comprised four distinct VLMs, each representing different approaches to visual-language understanding:

\begin{itemize}
    \item \textbf{GPT-4o}, developed by \textit{OpenAI}, is an integrated visual and textual processing through a unified transformer architecture with attention mechanisms \cite{openai2024gpt4technicalreport}. 
    \item \textbf{Claude 3.5 Sonnet} is an AI assistant developed by \textit{Anthropic} and employs constitutional AI principles to enhance reliability and minimize hallucinations \cite{anthropic}. 
    \item \textbf{Gemini 1.5 Pro}, previously known as \textit{Bard} and developed by \textit{Google DeepMind}, features end-to-end training on diverse visual-textual datasets, optimizing multimodal understanding \cite{GoogleDeepMind}. 
    \item \textbf{Llama3.2-vision} is an open-source architecture with transparent visual processing capabilities developed by \textit{Meta}. It offers models like the 11B and 90B variants that support image reasoning tasks such as document-level understanding, captioning, and visual grounding \cite{Llama32}. In this paper, we utilized the Llama 3.2 Vision 11B model. 
\end{itemize}
We standardized several key configuration parameters across all models to ensure consistent and reproducible results. We set the \textit{temperature} to 0, which minimizes response randomness by always selecting the most probable output tokens. This setting eliminates stochastic variation, making direct model comparisons more reliable. While alternative approaches—such as varying temperature settings—could provide insights into response variability, our goal was to establish a stable, deterministic benchmark. The \textit{max\_tokens} parameter was set to 300, limiting response length while ensuring sufficient detail for reasoning explanations. These settings prioritize deterministic behavior for our evaluation.

\subsection{Assessment Framework}
Our evaluation employs VLAT and CALVI, selected over alternatives like Mini-VLAT \cite{pandey2023mini} and specialized tests \cite{firat2020treemap, firat2022p, cui2023adaptive} for their comprehensive coverage of visualization tasks and chart types. 
\begin{itemize}
    \item \textbf{Visualization Literacy Assessment Test (VLAT)} provides a rigorous evaluation framework based on Classical Test Theory (CTT) \cite{devellis2006classical}, which analyzes item performance through basic statistics like item difficulty and discrimination indices, and it comprises of 53 multiple-choice items across 12 visualization types. Originally validated with 200 participants, it includes an ``Omit'' option and employs a correction-for-guessing formula (Equation~\ref{eq:corrected_score}) \cite{diamond1973correction, frary1988formula} to adjust scores based on incorrect responses. The test's complete-item approach enables direct comparison between human and VLM performance, with human scores typically ranging from 10.05 to 43.67 (\textit{M}=28.82, \textit{SD}=8.16) on the corrected scoring scale. 
    \item \textbf{Critical thinking Assessment for Literacy in Visualizations (CALVI)} employs Item Response Theory (IRT) \cite{embretson2013item}, which models the probability of correct responses based on both item properties and test-taker abilities to assess critical thinking through 45 items targeting various misleader types. The test consists of ``trick'' items using misleading and erroneous visualizations and ``normal'' items using well-formed visualizations inspired by VLAT. In its original validation study with 497 participants, each participant completed 30 items - 15 trick items randomly sampled from the bank of 45 items and 15 fixed normal items. The normal items serve as a baseline for assessing basic visualization interpretation abilities. Our study implements CALVI differently by having VLMs complete the full set of 45 trick items to enable comprehensive evaluation. In its original validation with human participants, performance on trick items ranged from 0\% to 93\% (\textit{M}=39\%, \textit{SD}=16\%). While this sampling approach complicates direct human-VLM comparisons, CALVI's systematic coverage of visualization deception makes it invaluable for assessing VLMs' critical analysis capabilities.
\end{itemize}
We chose these assessments' complementary strengths in evaluating different aspects of visualization literacy. Together, they provide a comprehensive framework for assessing both basic visualization interpretation skills and critical thinking abilities in the context of potentially misleading visualizations.

\begin{figure*}
    \centering
    \includegraphics[height=1.9in, width=0.48\textwidth]{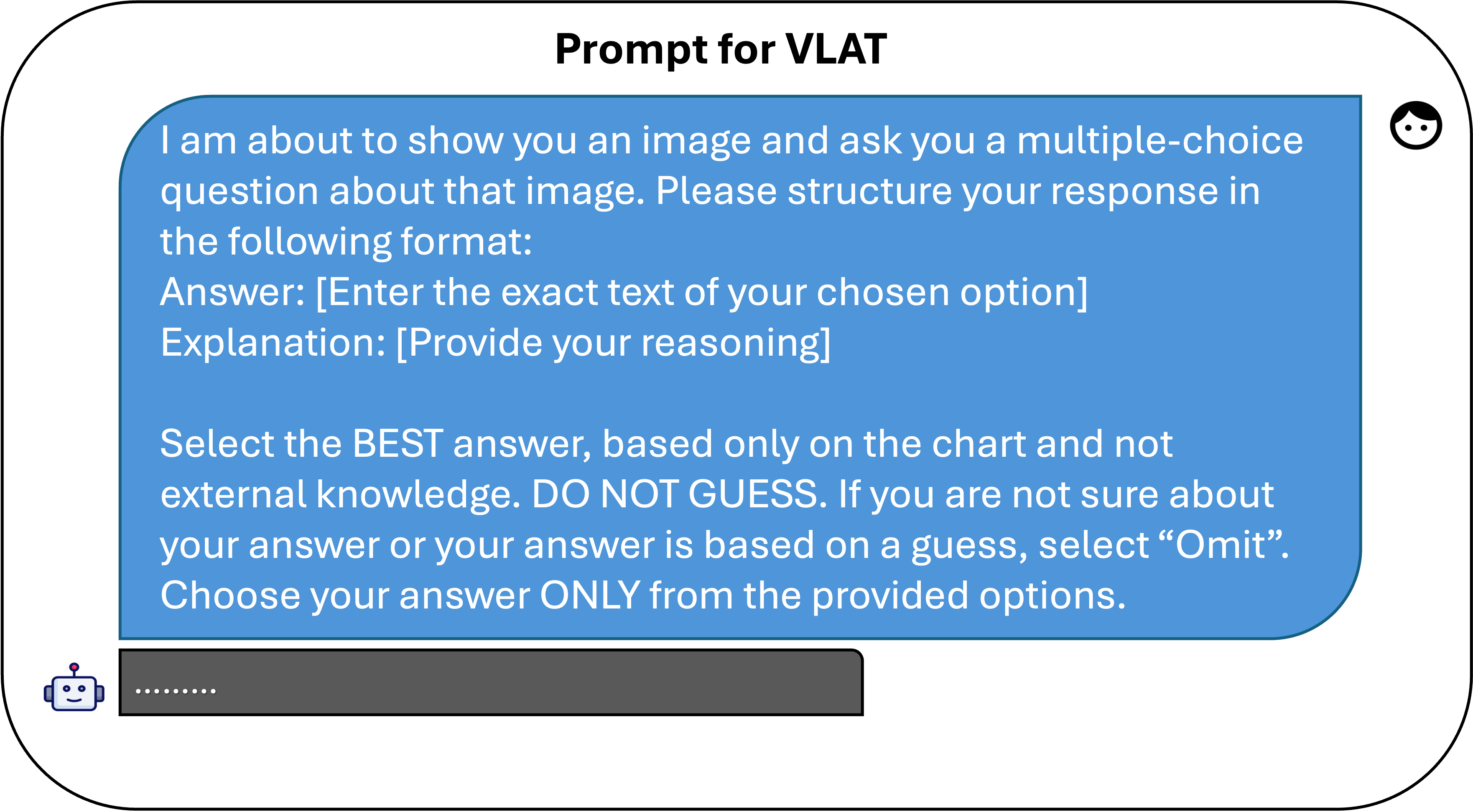}
    \quad
    \includegraphics[height=1.9in, width=0.48\textwidth]{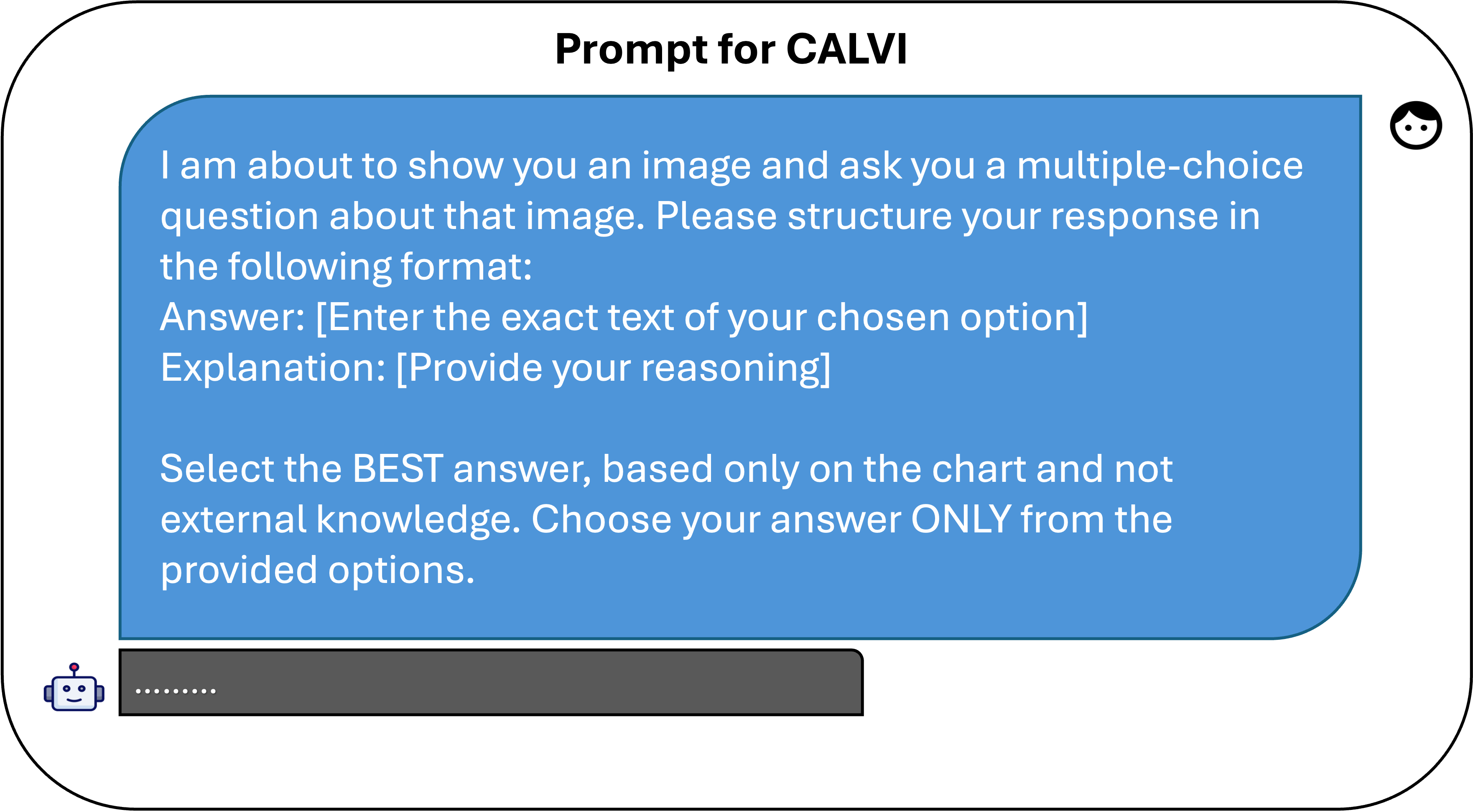}
    \caption{VLAT (left) and CALVI (right) prompt templates used for VLM evaluation.}
    \label{fig:prompts}
\end{figure*}

\subsection{Prompt Engineering and Design}
During prompt development, we began with a simple format: \textit{``Please select the correct option(s) from the given choices. Respond with the chosen option number(s) followed by `Why:' and then your explanation.''} However, pilot testing revealed two key limitations: VLMs did not choose the ``Omit'' option unless explicitly prompted, and their unstructured responses made it challenging to extract choices and explanations systematically. These findings led to the development of more structured prompts, building upon work by Bendeck et al. \cite{bendeck2024empirical}, who evaluated GPT-4's visualization literacy capabilities.

As shown in Figure \ref{fig:prompts}, we designed standardized prompts for both VLAT and CALVI assessments. The VLAT prompt explicitly instructs models to select answers based solely on chart information while discouraging guessing, with clear instructions about using the ``Omit'' option for uncertain responses. The CALVI~\cite{ge2023calvi} prompt follows a similar format, emphasizing response organization and basing decisions purely on chart information.

These prompts were designed to elicit structured responses that facilitate systematic analysis while maintaining consistency with the original assessment frameworks. The explicit formatting instructions ensure that model responses can be automatically processed and evaluated across multiple runs. Additionally, the prompts emphasize the importance of basing answers solely on the provided visualizations, helping to isolate the models' visualization literacy capabilities from their broader knowledge base.

\subsection{Evaluation Protocol} 
Each question is evaluated 10 times per model to account for potential variations in model responses. The questions are shown one at a time to prevent context contamination and to control the API rate limits. This also mitigates the potential interference between diverse visualization tasks while keeping consistent testing conditions across all models. The response collection follows a structured format, recording final answers and explanatory reasoning. For \VLAT, we track the use of the ``Omit'' option as an indicator of model uncertainty. This data collection enables detailed analysis of model performance and decision-making patterns. 

\section{Visualization Literacy Assessment Results}
\begin{table}[b!]
    \centering
    \begin{adjustbox}{max width=0.6\textwidth}
    \begin{tabular}{lcccccc}
        \toprule
        \textbf{Model} & \multicolumn{2}{c}{\textbf{VLAT (\%)}} & \multicolumn{2}{c}{\textbf{CALVI (\%)}} \\
        & \textbf{Mean} & \textbf{Std. Dev.} & \textbf{Mean} & \textbf{Std. Dev.} \\
        \midrule
        Human & 65.5 & 13.3 & 39.0 & 16 \\
        \midrule
        \Claude & 67.9 & 1.5 & 21.8 & 3.9 \\
        \GPT & 49.8 & 3.3 & 28.2 & 1.8 \\
        \Gemini & 42.5 & 3.0 & 30.0 & 4.4 \\
        \Llama & 43.8 & 3.6 & 24.4 & 6.0 \\
        \bottomrule
    \end{tabular}
    \end{adjustbox}
    \caption{Average model performance and standard deviation across 10 runs compared to human baseline, under the Random condition where both questions and answer options were randomized.}
    \label{tab:model-performance}
\end{table}
Our evaluation results in Table~\ref{tab:model-performance} represent averaged performance across 10 independent runs for each model on both \VLAT\ and \CALVI\ assessment under the Random condition. We report average across 10 runs to provide stable performance estimates. This extends beyond prior work by Bendeck et al. \cite{bendeck2024empirical}, which used 3 runs. Our pilot studies comparing 5 versus 10 runs revealed significant variance in model outputs, justifying our choice of 10 runs for more reliable measurements. The mean values with standard deviation offer insights into both performance level and consistency, enabling meaning comparisons with prior works. This repeated evaluation approach is crucial because VLMs can produce varying responses to the same question across runs, even with temperature set to 0. 

We first examine the corrected scores of both human and VLM performance on \VLAT. The corrected score (\(CS\)) for each model was computed using the formula established by Lee et al.~\cite{lee2016vlat}:
\begin{equation}
CS = R - \frac{W}{C-1}
\label{eq:corrected_score}
\end{equation}

where \(R\) represents the raw score (correct answers), \(W\) indicates incorrect answers, and \(C\) denotes number of choices available for each item. This formula adjusts for guessing by penalizing incorrect responses based on the number of possible choices. As shown in Table~\ref{tab:vlat-comparison}, our analysis encompasses 10 independent evaluation runs for each VLM, providing robust performance metrics compared against the original \VLAT\ study's 191 human participants. Human participants achieved higher corrected scores (\textit{M}=28.82) than most VLMs, though \Claude\ approached human-level performance with a corrected score of 28.96. Notably, VLMs showed consistent performance across runs, evidenced by lower standard deviations (0.77-2.67) than humans (8.16).

\begin{table}
    \centering
    \caption{Comparison of Humans and VLMs on VLAT}
    \begin{adjustbox}{max width=0.50\textwidth}
    \begin{tabular}{lccccc}
        \toprule
        \textbf{Model} & \textbf{Score Type} & \textbf{Mean (\textit{M})} & \textbf{Range} & \textbf{SD} \\
        \midrule
        Human \cite{lee2016vlat}          & Regular   & 34.72 & (14, 50)   & 7.05 \\
                      & Corrected & 28.82 & (10.05, 43.67)  & 8.16 \\
        \midrule
        \Claude\         & Regular   & 36.00 & (35, 37) & 0.77 \\
                      & Corrected & 28.96 & (27.55, 30.38)  & 1.10 \\
        \midrule
        \GPT\            & Regular   & 26.40 & (23, 29)      & 1.74 \\
                      & Corrected & 15.39 & (10.58, 19.06)  & 2.47 \\
        \midrule
        \Gemini\         & Regular   & 22.50 & (20, 26)      & 1.57 \\
                      & Corrected &  9.87 & (6.34, 14.82)   & 2.21 \\
        \midrule
        \Llama\          & Regular   & 23.20 & (20, 26)  & 1.89 \\
                      & Corrected & 10.86 & (6.34, 14.82) & 2.67 \\
        \bottomrule
    \end{tabular}
    \end{adjustbox}
    \label{tab:vlat-comparison}
\end{table}

The results reveal variations in VLMs' ability to understand and interpret visualizations across both assessments. \Claude\ demonstrates superior performance in \VLAT\ with an accuracy of 67.9\%, outperforming other models (\GPT: 49.8\%, \Gemini: 42.5\%, \Llama: 43.8\%). However, all models show notably lower performance on \CALVI, with accuracies ranging from 21.8\% (\Claude) to 30.0\% (\Gemini). This considerable performance gap between \VLAT\ and \CALVI\ suggests that while VLMs have developed reasonable capabilities for basic visualization interpretation tasks, they struggle significantly with identifying and reasoning about misleading elements in visualizations. 

\begin{figure*}
    \centering
        \includegraphics[width=0.95\textwidth]{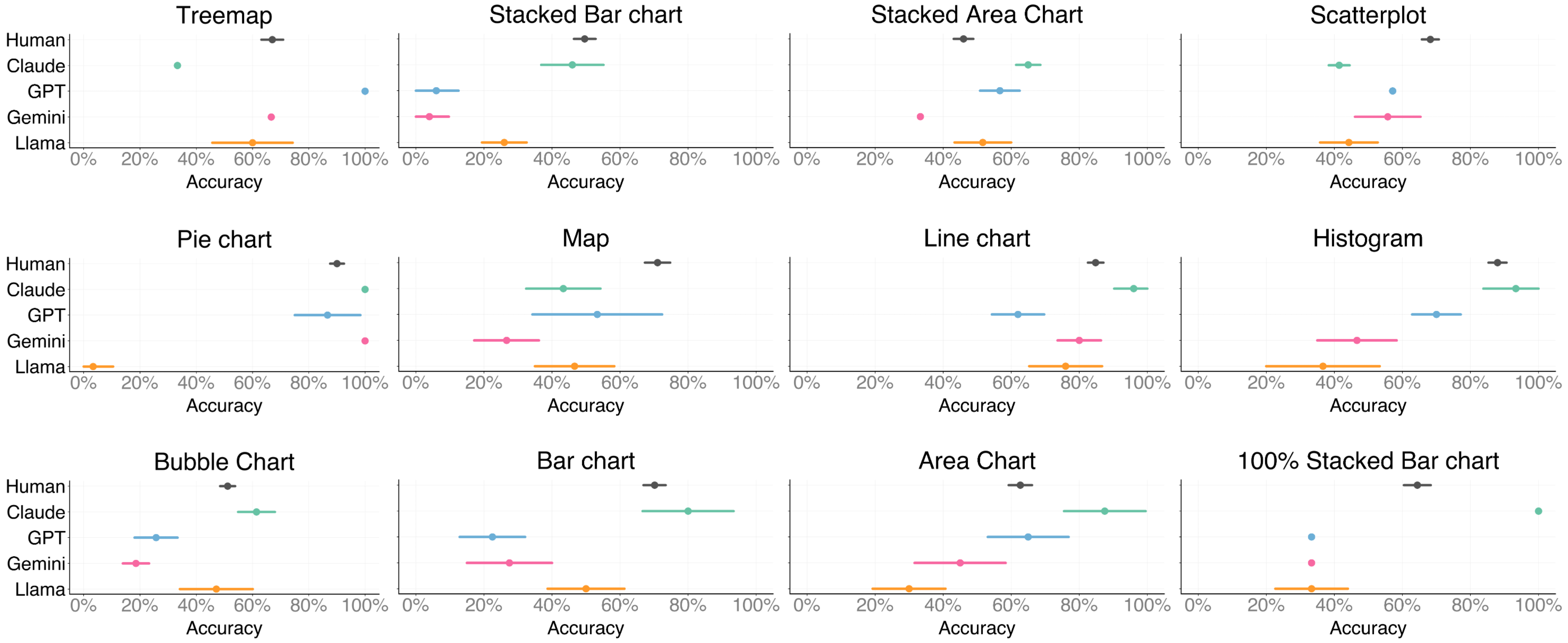}
    \caption{Model performance comparison across different visualization types in \VLAT. Each plot shows mean accuracy with 95\% confidence intervals, comparing VLMs against human performance.}
    \label{fig:chartype}
\end{figure*}

\begin{figure*}
    \centering
    \includegraphics[width=0.95\textwidth]{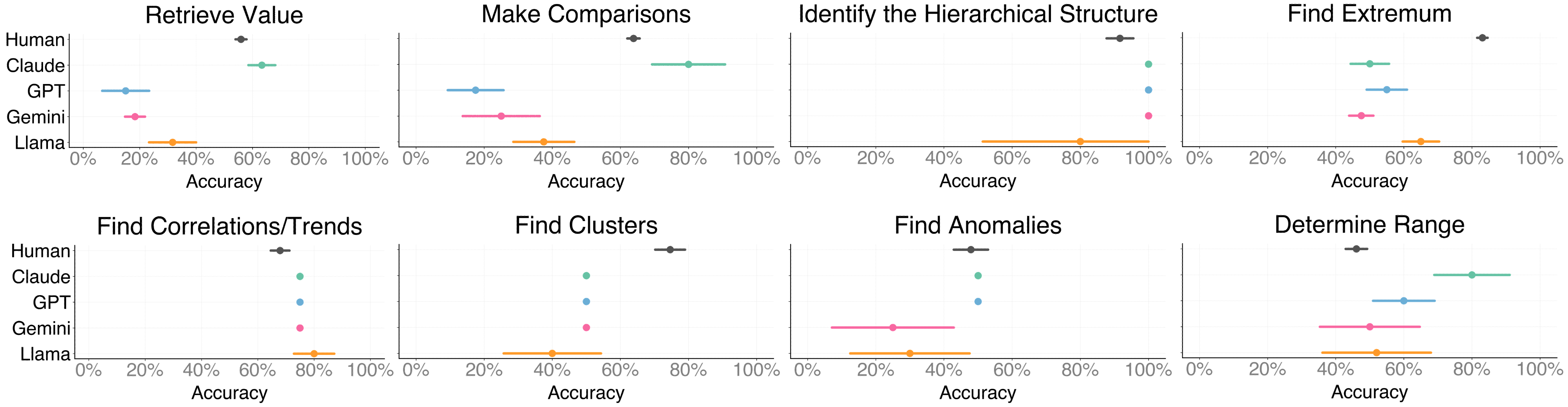}
    \caption{Model performance comparison across different visualization tasks in \VLAT. Each plot shows mean accuracy with 95\% confidence intervals, comparing VLMs against human performance.}
    \label{fig:task-comparison}
\end{figure*}

\subsection{Performance Across Chart Types}
Model performance analysis across different chart types reveals distinct patterns in VLMs' visualization interpretation capabilities and their relationship to human performance, as shown in Figure~\ref{fig:chartype}. While humans maintain relatively consistent performance across visualization types, VLMs exhibit considerable variability. \Claude\ demonstrates superior performance across most chart types, achieving perfect accuracy (100\%) on both 100\% stacked bar charts and pie charts (compared to human performance of 80-90\%), with exceptional performance on histograms (93.3\%) and line charts (96.0\%). All models show strong capabilities in line charts (\Claude: 96.0\%, \Gemini: 80.0\%, \GPT: 62.0\%, \Llama: 76.0\%). 

However, performance degrades significantly with visualization complexity, particularly with bubble charts encoding multiple variables simultaneously (accuracy range: 18.6-61.4\%). An interesting pattern emerges in spatial visualization interpretation: while models perform well on treemaps (\GPT: 100\%, \Gemini: 66.7\%, \Llama: 60.0\%), they struggle with map-based visualizations (26.7-53.3\%) compared to more consistent human performance. This disparity suggests stronger capabilities in processing hierarchical structures over geographic relationships. Basic visualization types like area charts and scatterplots show notable performance variation, highlighting inconsistencies in visual processing compared to the more uniform human performance patterns. 

These patterns, i.e., excelling in structural analysis while struggling with precise numerical tasks and complex spatial relationships, align with broader findings in VLM research. Alnegheimish et al. \cite{alnegheimish2024large} found that deep learning models outperform VLMs by approximately 30\% in anomaly detection. In contrast, Liu et al. \cite{liu2024large} demonstrated that VLMs require strategies like knowledge distillation for effective time series analysis.

\subsection{Task-Specific Performance}
The analysis of task-specific performance reveals nuanced patterns in how VLMs handle different visualization challenges, particularly compared to human benchmarks. Figure~\ref{fig:task-comparison} gives an overview of these patterns on various visualization tasks and highlights surprising strengths of the VLM capabilities along with significant weaknesses.

VLMs have shown great ability in pattern recognition and understanding of structure in tasks. Hierarchical structure identification is a particular strength, with \Claude, \Gemini, and \GPT\ achieving perfect accuracy (100\%) and \Llama\ maintaining strong performance (80\%), all above typical human performance of 90\%. Similarly, the models show consistent strength in finding correlations and trends at about 75-80\%, which closely approaches human levels of 70\%. This high performance in pattern-based tasks suggests that VLMs have developed robust capabilities in recognizing and interpreting systematic relationships within visualizations.

However, the performance terrain changes dramatically in tasks requiring precise numerical knowledge or more complex analytical reasoning. An analysis of value retrieval and evaluative tasks reveals significant disparities between models: while \Claude\ performs exceptionally well ($\approx$ 71\%), even surpassing standard human performance, other models demonstrate considerably lower accuracy (\Llama: 22.3\%, \Gemini: 37.1\%). This performance gap suggests that reasonable numerical interpretation is not inherent in VLMs but instead largely determined by specific architectural features or training methods. 

Anomaly detection presents another revealing challenge, with performance ranging from 25\% (\Gemini) to 50\% (\Claude), compared to consistent human performance around 50\%. This task, requiring both pattern recognition and deviation identification, seems to push the limits of current VLM capabilities. Interestingly, on range determination tasks, some VLMs even outperform humans, with \Claude\ achieving 80-90\% accuracy, whereas human performance is 45\%. This unexpected superiority in some quantitative tasks indicates possible advantages of computational approaches in specific analytical contexts.  

The most pronounced human-VLM performance differences appear in tasks requiring the integration of multiple visual elements or precise numerical comprehension. While humans maintain relatively consistent performance across diverse task types, VLMs show marked variability, particularly in tasks demanding detailed analysis or complex inference. These findings align with broader research on VLM limitations in numerical reasoning and anomaly detection \cite{alnegheimish2024large}, suggesting the need for specialized strategies like knowledge distillation to enhance performance \cite{liu2024large}. The observed patterns point to fundamental architectural limitations in processing complex numerical relationships, providing clear directions for future VLM development.

\subsection{Uncertainty Analysis}
Adding an ``Omit'' option in \VLAT\ brings rich insights into how VLMs handle uncertainty in visualization interpretation tasks. Models exhibit distinct patterns in their uncertainty expression, as shown in Table~\ref{tab:omit-counts}. \Gemini\ has the highest propensity to acknowledge uncertainty, choosing to omit responses for 22.5\% of questions (averaging 11.9 omissions per run). This conservative approach reflects a greater awareness of uncertainty compared to other models.
\begin{table}[h!]
    \centering
    \caption{``Omit'' response frequency by model across 10 runs, shown as counts and percentages of total VLAT questions.}
    \begin{adjustbox}{max width=0.50\textwidth}
    \begin{tabular}{lccc}
        \toprule
        \textbf{Model} & \textbf{Average Omits} & \textbf{Range} & \textbf{Percentage (\%)} \\
        \midrule
        \Claude\ & 4.3 & 4--6 & 8.1 \\
        \GPT\ & 3.8 & 2--6 & 7.2 \\
        \Gemini\ & 11.9 & 11--14 & 22.5 \\
        \Llama\ & 4.3 & 3--6 & 8.1 \\
        \bottomrule
    \end{tabular}
    \end{adjustbox}
    \label{tab:omit-counts}
\end{table}

\begin{figure*}[t!]
    \centering
    \includegraphics[width=0.95\linewidth]{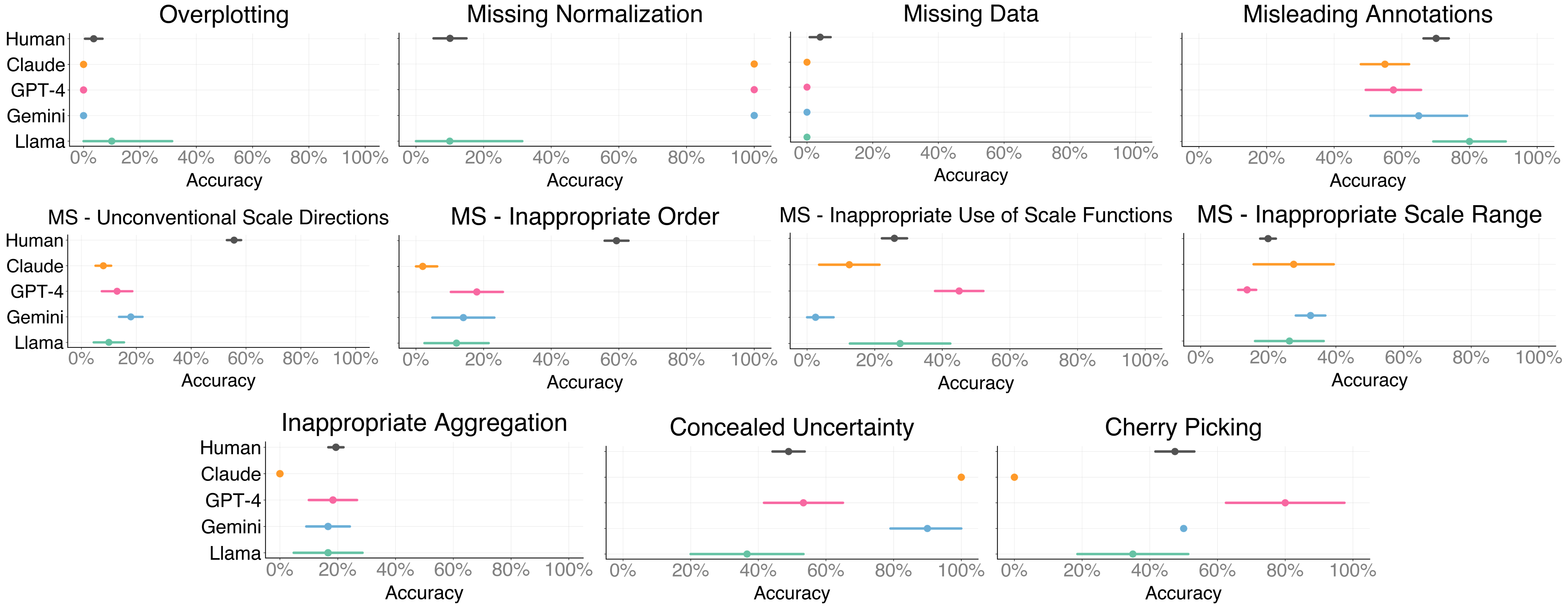}
    \caption{Comparison of VLM and human performance across different misleader types in \CALVI\ assessment. Points show mean accuracy with 95\% confidence intervals. MS = Manipulation of Scales.}
    \label{fig:calvi-misleader}
\end{figure*}

\section{Critical Thinking Assessment Results}
A comparison of model performance across \CALVI's misleader types reveals systematic patterns in VLMs' ability to detect visualization deceptions, as shown in Figure~\ref{fig:calvi-misleader}. All models detect missing normalization errors with 100\% accuracy (except \Llama\ at 10\%). Models also demonstrate strong capabilities in detecting concealed uncertainty, with \Claude\ and \Gemini\ showing particularly high performance (100\% and 90\%, respectively). Interestingly, models show varied capabilities in specific misleader categories. \GPT\ is good at detecting cherry picking (80\%), while \Llama\ shows superior performance in identifying misleading annotations (80\%). However, all models struggle significantly with certain types of misleaders:
\begin{itemize}
\item \textbf{Missing Data}: All models consistently fail to identify missing data issues (0\% accuracy across all models). 
\item \textbf{Scale Manipulations}: Performance is notably poor when dealing with inappropriate scale orders (12-18\%) and inappropriate scale ranges (13.8-32.5\%).
\item \textbf{Overplotting}: Most models (\Claude, \GPT, \Gemini) completely fail to detect overplotting issues (0\% accuracy), with only \Llama\ showing minimal capability (10\%). 
\end{itemize}
VLMs show an intriguing blind spot: while they excel at catching obvious chart errors like missing labels, they're surprisingly weak at spotting subtler forms of visual deception.

\subsection{Model-Specific Capabilities}
Analysis of the models reveals distinct patterns in their ability to detect specific visualization misleaders. \Claude\ demonstrates exceptional capabilities in identifying statistical and methodological flaws, achieving perfect accuracy (100\%) in both concealed uncertainty and missing normalization categories. \GPT\ exhibits particular strength in detecting cherry picking (80\% accuracy), suggesting advanced capabilities in identifying selective data presentation bias. \Gemini\ shows balanced performance across categories with notable excellence in concealed uncertainty (90\%) and missing normalization detection (100\%). \Llama\ , despite the lower overall performance, shows specialized expertise in detecting misleading annotations (80\% accuracy), indicating potential in a contextual analysis of visualization elements. These diverse performance profiles suggest that architectural and training differences among models may foster specialized capabilities in detecting specific types of visualization deception. \Claude\ 's perfect accuracy in uncertainty detection likely stems from robust statistical reasoning capabilities, while \GPT\ 's proficiency in identifying cherry-picking suggests advanced pattern recognition mechanisms. 

\subsection{Comparative Analysis with Human Performance}
The comparative analysis of \CALVI\ performance presents distinct methodological considerations that affect human-VLM comparisons. While our VLM assessment utilized CALVI's complete set of 45 misleader questions (achieving 21.8\%-30.0\% accuracy), the original human study followed a different protocol - approximately 500 participants each completed 30 items total: 15 trick items randomly sampled from the bank of 45 items (achieving \textit{M}=39\%, \textit{SD}=16\%) and 15 fixed normal items (achieving \textit{M}=80\%, \textit{SD}=13\%). This methodological distinction contrasts VLAT's approach, where all participants completed the whole set. 

Analysis of performance across misleader types (Figure~\ref{fig:calvi-misleader}) reveals intriguing patterns in deception detection capabilities. VLMs show remarkable proficiency in certain areas, with \Claude, \GPT, and \Gemini\ achieving perfect accuracy (100\%) in missing normalization, significantly outperforming humans (10\%). Similarly, for concealed uncertainty, \Claude\ and \Gemini\ demonstrate exceptional capabilities (100\% and 90\% respectively) compared to humans ($\approx$ 50\%).

However, VLMs struggle with subtle manipulations. In detecting unconventional scale directions, humans maintain approximately 50\% accuracy while VLMs perform poorly (8-18\%). The challenges extend to missing data detection, where all models show 0\% accuracy, and overplotting identification, where most VLMs fail. These limitations align with known constraints in processing fine-grained visual information \cite{islam2024large} and numerical reasoning \cite{zou2024dynamath}.

Some misleader types reveal more nuanced patterns. In detecting misleading annotations, VLMs achieve varying success (50-80\%), approaching human performance levels. \GPT shows particular strength in cherry-picking detection (80\% accuracy), exceeding human performance. These findings, supported by research on VLM capabilities \cite{nagar2024zero}, suggest that while these models can effectively identify obvious visualization errors, they struggle with sophisticated forms of visual deception, requiring an integrated understanding of visualization principles and contextual reasoning. This performance gap emphasizes the continued importance of human oversight in critical visualization analysis.

\subsection{Common Failure Patterns}
Analysis across all models reveals systematic weaknesses in visualization deception detection, highlighting fundamental limitations in current VLM architectures. Most critically, all four VLMs demonstrate complete failure (0\% accuracy) in identifying missing data issues, indicating a universal inability to detect omitted information in visualizations. 

Scale manipulation detection emerges as another significant challenge. Models consistently struggle with identifying inappropriate scale ordering and ranges, with accuracy varying from 2.0\% to 32.5\%. This persistent weakness suggests fundamental limitations in evaluating visualization scaling decisions. Similarly, complex visualization issues, particularly overplotting detection, pose substantial challenges, with three models showing complete failure and \Llama\ achieving minimal success (10\% accuracy).

Function-related deceptions represent another area of universal difficulty, evidenced by consistently poor performance in detecting inappropriate scale functions (2.5 - 45.0\%) and scale function manipulation (8.0 - 18.0\%). These patterns point to three fundamental limitations in current VLM architectures: inadequate detail-oriented analysis capabilities \cite{nagar2024zero}, limited mathematical reasoning \cite{zou2024dynamath}, and weak structural analysis skills \cite{islam2024large}. These findings underscore critical areas requiring enhancement in future VLM development, particularly in strengthening capabilities for detailed visual analysis and mathematical relationship comprehension in data visualization.

\section{Discussion}
Our systematic evaluation of VLMs' visualization literacy offers insights for visualization researchers and practitioners. These findings illuminate the potential and limitations of current VLM architectures, providing a roadmap for future research and practical applications in visualization systems. Below, we expand on the key results, contextualize them within prior work, and propose avenues for development.

\subsection{Comparing Claude's Visualization Literacy to Human Performance}
Among the models we evaluated, \Claude\ demonstrated human-comparable performance on \VLAT, marking a significant milestone in VLM capabilities. With an overall accuracy of 67.9\%, \Claude\ outperformed not only other VLMs but also approached human-level performance in several visualization types, including stacked area charts (85\% vs 70\% human accuracy), histograms (93.3\% vs 75\%), and line charts (96\% vs 85\%). However, without transparency regarding \Claude's architecture, attributing this success to specific technical innovations remains challenging.

Despite strong overall performance, \Claude\ showed consistent weaknesses in spatial encoding tasks (maps, scatterplots, treemaps), suggesting fundamental limitations in spatial reasoning common to current VLM architectures. This pattern indicates a hierarchical nature in VLMs' visualization literacy: strong performance in pattern recognition (75-80\% accuracy in trend identification) but degrading capabilities as tasks require more complex reasoning or precise numerical analysis. Performance particularly declined in tasks requiring multiple variable interpretation, such as bubble charts (18.6-61.4\% accuracy range across models).

This “complexity threshold” within current architectures—whether due to attention mechanisms, spatial encoding limitations, or training data—presents a crucial area for future research. Understanding these limitations could guide the development of hybrid models or specialized training approaches that better handle complex visualization tasks while maintaining the strong performance in basic pattern recognition.

\subsection{Uncertainty Management Strategies}
Our analysis reveals distinct patterns in how VLMs manage uncertainty during visualization tasks. Most notably, \Gemini’s omission rate of 22.5\% reflects a conservative approach to uncertainty, reducing false positions but potentially limiting insights in exploratory analysis where incomplete answers have value. This finding aligns with research on uncertainty quantification in AI \cite{ji2023survey} and suggests the need for models with dynamic risk tolerance adjustable to specific tasks and contexts. Future visualization systems could benefit from interfaces allowing users to calibrate model uncertainty thresholds based on their application needs. 

\subsection{Poor Performance in Detecting Misleading Visualizations}
All models, including \Claude, demonstrated limited capabilities in identifying misleading visualizations, with \CALVI\ accuracies ranging from 21.8\% to 30.0\%. This performance gap is particularly notable in detecting subtle manipulations such as inappropriate scale ordering (8-18\% accuracy) and overplotting (0-10\% accuracy). These findings contrast with prior work where VLMs showed better performance when explicitly prompted to evaluate misleading elements \cite{lo2024good}, suggesting that task framing significantly influences performance. 

The models showed striking disparities in their detection capabilities: while achieving near-perfect accuracy in identifying missing normalization (100\% for most models) and concealed uncertainty (90-100\% for top performers), they universally failed at detecting missing data (0\% accuracy) and struggled with scale manipulations (13.8-32.5\% accuracy). This pattern suggests that current VLMs excel at identifying obvious structural issues but struggle with more nuanced forms of visualization deception, highlighting a critical gap in their analytical capabilities that requires attention in future development.

\subsection{VLM Strengths With Utility for Visualization Systems}
The strong pattern recognition of VLMs make them valuable tools for initial data exploration and basic chart interpretation. Their high accuracy in trend identification (75-80\%) and hierarchical structure detection (80-100\%) suggests applications in educational settings, data journalism, or preliminary data analysis. However, their limitations in detecting misleading visualizations necessitate careful integration into practical applications.

We recommend a hybrid approach where VLMs handle initial analysis tasks while maintaining human oversight for critical decisions. Such an approach would employ VLMs for rapid initial pattern detection and trend summarization, while implementing confidence thresholds that trigger human review for complex or potentially misleading visualizations. Additionally, visualization systems should develop interfaces that clearly communicate model uncertainty and limitations to users. This strategy would leverage VLMs' computational efficiency while maintaining the critical thinking and context awareness that human analysts provide, particularly in scenarios where visualization misinterpretation could have significant consequences.

\section{Limitations}
Our study provides important insights into VLM visualization literacy capabilities, but several important limitations must be considered. The fundamental limitations begin with our assessment methodology. Although \VLAT\ and \CALVI\ are well-validated assessment instruments in the field of visualization literacy, their multiple-choice format may not capture the full range of VLM capabilities, potentially limiting the expression of more sophisticated reasoning processes \cite{ge2023calvi, brown2020language}. These assessments also cannot evaluate important aspects of modern visualization practices, such as interactive visualization interpretation or dynamic data representations \cite{heer2012interactive}. 

Our choices in model selection and configuration introduce another layer of constraints. We evaluated four prominent VLMs under specific parameter settings (temperature = 0), privileging consistency and reproducibility. However, this approach may not bring out the potential of these models under more flexible configurations, such as those employing stochastic sampling or incorporating human feedback mechanisms \cite{openai2024gpt4technicalreport, touvron2023llama}. Given the rapid development in VLM technology, newer architectures may mitigate some limitations we identified in visual processing capabilities.\\

A more fundamental challenge stems from the intrinsic differences between human and VLM approaches to visualization processing. While human analysts naturally incorporate domain expertise and holistic reasoning strategies, VLMs operate within more constrained computational frameworks \cite{bubeck2023sparks}. This cognitive gap is particularly evident in \CALVI\ tasks, where identifying subtle misleading elements requires critical thinking abilities that current VLM architectures struggle to replicate.

\section{Future Work}
Our findings reveal several promising research directions for advancing VLM capabilities in visualization interpretation: 

\textbf{Fine-tuning VLMs for Visualization Literacy} Fine-tuning models on visualization-centric datasets could bridge the observed gaps in spatial reasoning and quantitative comprehension. The datasets should further include various types of visualizations, such as deceptive designs like overplotting, skewed baselines, and misleading scales, that would eventually train the model to recognize and respond to misleaders more precisely \cite{pandey2015deceptive, lo2022misinformed}. Techniques such as reinforcement learning with human feedback (RLHF) or curriculum learning could allow progressive improvements in handling complex visualizations \cite{ouyang2022training}.

\textbf{Leveraging Human-AI Synergy} While VLMs exhibit clear strengths in pattern recognition and trend analysis, their weaknesses in tasks requiring numerical precision or critical evaluation necessitate hybrid frameworks. Such systems could use VLMs for preliminary tasks, such as identifying patterns or anomalies, and rely on human oversight for high-stakes decisions, such as identifying misleaders. This collaborative approach could leverage the respective strengths of both humans and VLMs to improve accuracy and reliability in visualization tasks \cite{wang2024human}.

\textbf{Advancing Architectural Innovations} Architectural advancements, such as vision transformers and multi-modal learning approaches, hold promise for addressing the limitations identified in our study \cite{liu2023survey}. Another promising avenue is the exploration of uncertainty quantification methods to make VLMs more robust in ambiguous or incomplete visualizations, as shown by recent research on probabilistic model outputs \cite{abdar2021review}.

\textbf{Enhanced Prompt Engineering Strategies} Our findings highlight how task framing impacts VLM performance in visualization analysis. Future research should explore sophisticated prompting strategies, such as chain-of-thought prompts or hierarchical task decompositions, to enhance VLMs' critical analysis \cite{wei2022chain}. This could include developing standardized prompt templates for different visualization tasks and investigating how varying levels of explicit instruction affect model performance in detecting misleading elements. While our study used standardized prompts to ensure fair comparisons, future work could explore more advanced strategies like multi-shot prompting, chain-of-thought reasoning, and model-specific prompt optimization \cite{jeong2024limited}. These tailored approaches may provide deeper insights into model-specific strengths and enhance visualization literacy.

\section{Conclusion}
Our comprehensive evaluation of VLMs' visualization literacy capabilities reveals a complex landscape of advances and limitations. While some models approach human-level performance in basic visualization tasks (\Claude\ achieving 67.9\% on \VLAT), they struggle significantly with critical thinking and detecting visualization deception (21.8-30.0\% on \CALVI). These stark performance variations demonstrate that current VLMs are best suited as assistive tools rather than autonomous systems. Our reproducible framework provides a foundation for future VLM evaluations while emphasizing the importance of balanced human-AI collaboration in visualization analysis. 

\section{Acknowledgments}
This project was partially supported by the National Science Foundation under OAC-2118201 and IIS-2142977. We thank Bum Chul Kwon and Sung-Hee Kim for sharing the data from the VLAT study. Lastly, we also thank the authors of CALVI for sharing the data in an open-source repository. 

\bibliographystyle{eg-alpha-doi} 
\bibliography{egbibsample}       

\newcommand{\etalchar}[1]{$^{#1}$}
\begin{thebibliography}{\uppercase{BGMMS21}}

\bibitem[ADL{\etalchar{*}}22]{alayrac2022flamingo}
\textsc{Alayrac J.-B., Donahue J., Luc P., Miech A., Barr I., Hasson Y., Lenc K., Mensch A., Millican K., Reynolds M., et~al.}:
\newblock Flamingo: a visual language model for few-shot learning.
\newblock \emph{Advances in neural information processing systems 35} (2022), 23716--23736.

\bibitem[ANBEV24]{alnegheimish2024large}
\textsc{Alnegheimish S., Nguyen L., Berti-Equille L., Veeramachaneni K.}:
\newblock Large language models can be zero-shot anomaly detectors for time series?
\newblock \emph{arXiv preprint arXiv:2405.14755} (2024).

\bibitem[Ant]{anthropic}
\textsc{Anthropic}:
\newblock \ claude 3.5 sonnet.
\newblock URL: \url{https://www.anthropic.com/claude/sonnet}.

\bibitem[APH{\etalchar{*}}21]{abdar2021review}
\textsc{Abdar M., Pourpanah F., Hussain S., Rezazadegan D., Liu L., Ghavamzadeh M., Fieguth P., Cao X., Khosravi A., Acharya U.~R., et~al.}:
\newblock A review of uncertainty quantification in deep learning: Techniques, applications and challenges.
\newblock \emph{Information fusion 76} (2021), 243--297.

\bibitem[BCE{\etalchar{*}}23]{bubeck2023sparks}
\textsc{Bubeck S., Chandrasekaran V., Eldan R., Gehrke J., Horvitz E., Kamar E., Lee P., Lee Y.~T., Li Y., Lundberg S., et~al.}:
\newblock Sparks of artificial general intelligence: Early experiments with gpt-4.
\newblock \emph{arXiv preprint arXiv:2303.12712} (2023).

\bibitem[BGMMS21]{bender2021dangers}
\textsc{Bender E.~M., Gebru T., McMillan-Major A., Shmitchell S.}:
\newblock On the dangers of stochastic parrots: Can language models be too big?
\newblock In \emph{Proceedings of the 2021 ACM conference on fairness, accountability, and transparency} (2021), pp.~610--623.

\bibitem[BMBH16]{borner2016investigating}
\textsc{B{\"o}rner K., Maltese A., Balliet R.~N., Heimlich J.}:
\newblock Investigating aspects of data visualization literacy using 20 information visualizations and 273 science museum visitors.
\newblock \emph{Information Visualization 15}, 3 (2016), 198--213.

\bibitem[BPA{\etalchar{*}}24]{bordes2024introduction}
\textsc{Bordes F., Pang R.~Y., Ajay A., Li A.~C., Bardes A., Petryk S., Ma{\~n}as O., Lin Z., Mahmoud A., Jayaraman B., et~al.}:
\newblock An introduction to vision-language modeling.
\newblock \emph{arXiv preprint arXiv:2405.17247} (2024).

\bibitem[BRBF14]{boy2014principled}
\textsc{Boy J., Rensink R.~A., Bertini E., Fekete J.-D.}:
\newblock A principled way of assessing visualization literacy.
\newblock \emph{IEEE transactions on visualization and computer graphics 20}, 12 (2014), 1963--1972.

\bibitem[Bro20]{brown2020language}
\textsc{Brown T.~B.}:
\newblock Language models are few-shot learners.
\newblock \emph{arXiv preprint arXiv:2005.14165} (2020).

\bibitem[BS24]{bendeck2024empirical}
\textsc{Bendeck A., Stasko J.}:
\newblock An empirical evaluation of the gpt-4 multimodal language model on visualization literacy tasks.
\newblock \emph{IEEE Transactions on Visualization and Computer Graphics} (2024).

\bibitem[CLD{\etalchar{*}}23]{cui2023adaptive}
\textsc{Cui Y., Lily W.~G., Ding Y., Yang F., Harrison L., Kay M.}:
\newblock Adaptive assessment of visualization literacy.
\newblock \emph{IEEE Transactions on Visualization and Computer Graphics} (2023).

\bibitem[CLL{\etalchar{*}}24]{choe2024enhancing}
\textsc{Choe K., Lee C., Lee S., Song J., Cho A., Kim N.~W., Seo J.}:
\newblock Enhancing data literacy on-demand: Llms as guides for novices in chart interpretation.
\newblock \emph{IEEE Transactions on Visualization and Computer Graphics} (2024).

\bibitem[CPG{\etalchar{*}}23]{chan2023clair}
\textsc{Chan D., Petryk S., Gonzalez J.~E., Darrell T., Canny J.}:
\newblock Clair: Evaluating image captions with large language models.
\newblock \emph{arXiv preprint arXiv:2310.12971} (2023).

\bibitem[CZX{\etalchar{*}}24]{chen2024viseval}
\textsc{Chen N., Zhang Y., Xu J., Ren K., Yang Y.}:
\newblock Viseval: A benchmark for data visualization in the era of large language models.
\newblock \emph{IEEE Transactions on Visualization and Computer Graphics} (2024).

\bibitem[DE73]{diamond1973correction}
\textsc{Diamond J., Evans W.}:
\newblock The correction for guessing.
\newblock \emph{Review of educational research 43}, 2 (1973), 181--191.

\bibitem[Dee]{GoogleDeepMind}
\textsc{DeepMind G.}:
\newblock Google pro.
\newblock URL: \url{https://deepmind.google/technologies/gemini/pro/}.

\bibitem[DeV06]{devellis2006classical}
\textsc{DeVellis R.~F.}:
\newblock Classical test theory.
\newblock \emph{Medical care 44}, 11 (2006), S50--S59.

\bibitem[ER13]{embretson2013item}
\textsc{Embretson S.~E., Reise S.~P.}:
\newblock \emph{Item response theory}.
\newblock Psychology Press, 2013.

\bibitem[FDL20]{firat2020treemap}
\textsc{Firat E., Denisova A., Laramee R.}:
\newblock Treemap literacy: A classroom-based investigation.
\newblock In \emph{Eurographics Proceedings} (2020).

\bibitem[FDWL22]{firat2022p}
\textsc{Firat E.~E., Denisova A., Wilson M.~L., Laramee R.~S.}:
\newblock P-lite: A study of parallel coordinate plot literacy.
\newblock \emph{Visual Informatics 6}, 3 (2022), 81--99.

\bibitem[Fra88]{frary1988formula}
\textsc{Frary R.~B.}:
\newblock Formula scoring of multiple-choice tests (correction for guessing).
\newblock \emph{Educational measurement: Issues and practice 7}, 2 (1988), 33--38.

\bibitem[GCK23]{ge2023calvi}
\textsc{Ge L.~W., Cui Y., Kay M.}:
\newblock Calvi: Critical thinking assessment for literacy in visualizations.
\newblock In \emph{Proceedings of the 2023 CHI conference on human factors in computing systems} (2023), pp.~1--18.

\bibitem[GKS{\etalchar{*}}24]{guo2024understanding}
\textsc{Guo G., Kang J.~J., Shah R.~S., Pfister H., Varma S.}:
\newblock Understanding graphical perception in data visualization through zero-shot prompting of vision-language models.
\newblock \emph{arXiv preprint arXiv:2411.00257} (2024).

\bibitem[GKWK24]{gorniak2024vizability}
\textsc{Gorniak J., Kim Y., Wei D., Kim N.~W.}:
\newblock Vizability: Enhancing chart accessibility with llm-based conversational interaction.
\newblock In \emph{Proceedings of the 37th Annual ACM Symposium on User Interface Software and Technology} (2024), pp.~1--19.

\bibitem[GLL{\etalchar{*}}23]{guo2023images}
\textsc{Guo J., Li J., Li D., Tiong A. M.~H., Li B., Tao D., Hoi S.}:
\newblock From images to textual prompts: Zero-shot visual question answering with frozen large language models.
\newblock In \emph{Proceedings of the IEEE/CVF conference on computer vision and pattern recognition} (2023), pp.~10867--10877.

\bibitem[HLL{\etalchar{*}}24]{hong2024data}
\textsc{Hong S., Lin Y., Liu B., Wu B., Li D., Chen J., Zhang J., Wang J., Zhang L., Zhuge M., et~al.}:
\newblock Data interpreter: An llm agent for data science.
\newblock \emph{arXiv preprint arXiv:2402.18679} (2024).

\bibitem[HS12]{heer2012interactive}
\textsc{Heer J., Shneiderman B.}:
\newblock Interactive dynamics for visual analysis: A taxonomy of tools that support the fluent and flexible use of visualizations.
\newblock \emph{Queue 10}, 2 (2012), 30--55.

\bibitem[HXL{\etalchar{*}}24]{hu2024bliva}
\textsc{Hu W., Xu Y., Li Y., Li W., Chen Z., Tu Z.}:
\newblock Bliva: A simple multimodal llm for better handling of text-rich visual questions.
\newblock In \emph{Proceedings of the AAAI Conference on Artificial Intelligence} (2024), vol.~38, pp.~2256--2264.

\bibitem[IRM{\etalchar{*}}24]{islam2024large}
\textsc{Islam M.~S., Rahman R., Masry A., Laskar M. T.~R., Nayeem M.~T., Hoque E.}:
\newblock Are large vision language models up to the challenge of chart comprehension and reasoning? an extensive investigation into the capabilities and limitations of lvlms.
\newblock \emph{arXiv preprint arXiv:2406.00257} (2024).

\bibitem[JLF{\etalchar{*}}23]{ji2023survey}
\textsc{Ji Z., Lee N., Frieske R., Yu T., Su D., Xu Y., Ishii E., Bang Y.~J., Madotto A., Fung P.}:
\newblock Survey of hallucination in natural language generation.
\newblock \emph{ACM Computing Surveys 55}, 12 (2023), 1--38.

\bibitem[JMG{\etalchar{*}}24]{jeong2024limited}
\textsc{Jeong D.~P., Mani P., Garg S., Lipton Z.~C., Oberst M.}:
\newblock The limited impact of medical adaptation of large language and vision-language models.
\newblock \emph{arXiv preprint arXiv:2411.08870} (2024).

\bibitem[KPR23]{kavaz2023chatbot}
\textsc{Kavaz E., Puig A., Rodr{\'\i}guez I.}:
\newblock Chatbot-based natural language interfaces for data visualisation: A scoping review.
\newblock \emph{Applied Sciences 13}, 12 (2023), 7025.

\bibitem[LBPL19]{lu2019vilbert}
\textsc{Lu J., Batra D., Parikh D., Lee S.}:
\newblock Vilbert: Pretraining task-agnostic visiolinguistic representations for vision-and-language tasks.
\newblock \emph{Advances in neural information processing systems 32} (2019).

\bibitem[LGS{\etalchar{*}}22]{lo2022misinformed}
\textsc{Lo L. Y.-H., Gupta A., Shigyo K., Wu A., Bertini E., Qu H.}:
\newblock Misinformed by visualization: What do we learn from misinformative visualizations?
\newblock In \emph{Computer Graphics Forum} (2022), vol.~41, Wiley Online Library, pp.~515--525.

\bibitem[LHZ{\etalchar{*}}24]{liu2024large}
\textsc{Liu C., He S., Zhou Q., Li S., Meng W.}:
\newblock Large language model guided knowledge distillation for time series anomaly detection.
\newblock \emph{arXiv preprint arXiv:2401.15123} (2024).

\bibitem[LKK16]{lee2016vlat}
\textsc{Lee S., Kim S.-H., Kwon B.~C.}:
\newblock Vlat: Development of a visualization literacy assessment test.
\newblock \emph{IEEE transactions on visualization and computer graphics 23}, 1 (2016), 551--560.

\bibitem[LLWL24]{liu2024visual}
\textsc{Liu H., Li C., Wu Q., Lee Y.~J.}:
\newblock Visual instruction tuning.
\newblock \emph{Advances in neural information processing systems 36} (2024).

\bibitem[LM22]{liew2022using}
\textsc{Liew A., Mueller K.}:
\newblock Using large language models to generate engaging captions for data visualizations.
\newblock \emph{arXiv preprint arXiv:2212.14047} (2022).

\bibitem[LMX{\etalchar{*}}22]{lu2022learn}
\textsc{Lu P., Mishra S., Xia T., Qiu L., Chang K.-W., Zhu S.-C., Tafjord O., Clark P., Kalyan A.}:
\newblock Learn to explain: Multimodal reasoning via thought chains for science question answering.
\newblock \emph{Advances in Neural Information Processing Systems 35} (2022), 2507--2521.

\bibitem[LQ24]{lo2024good}
\textsc{Lo L. Y.-H., Qu H.}:
\newblock How good (or bad) are llms at detecting misleading visualizations?
\newblock \emph{IEEE Transactions on Visualization and Computer Graphics} (2024).

\bibitem[LYT{\etalchar{*}}23]{liu2023trustworthy}
\textsc{Liu Y., Yao Y., Ton J.-F., Zhang X., Cheng R. G.~H., Klochkov Y., Taufiq M.~F., Li H.}:
\newblock Trustworthy llms: A survey and guideline for evaluating large language models' alignment.
\newblock \emph{arXiv preprint arXiv:2308.05374} (2023).

\bibitem[LYY{\etalchar{*}}19]{li2019visualbert}
\textsc{Li L.~H., Yatskar M., Yin D., Hsieh C.-J., Chang K.-W.}:
\newblock Visualbert: A simple and performant baseline for vision and language.
\newblock \emph{arXiv preprint arXiv:1908.03557} (2019).

\bibitem[LZW{\etalchar{*}}23]{liu2023survey}
\textsc{Liu Y., Zhang Y., Wang Y., Hou F., Yuan J., Tian J., Zhang Y., Shi Z., Fan J., He Z.}:
\newblock A survey of visual transformers.
\newblock \emph{IEEE Transactions on Neural Networks and Learning Systems} (2023).

\bibitem[MDW{\etalchar{*}}23]{ma2023insightpilot}
\textsc{Ma P., Ding R., Wang S., Han S., Zhang D.}:
\newblock Insightpilot: An llm-empowered automated data exploration system.
\newblock In \emph{Proceedings of the 2023 Conference on Empirical Methods in Natural Language Processing: System Demonstrations} (2023), pp.~346--352.

\bibitem[Met]{Llama32}
\textsc{Meta}:
\newblock Llama can now see and run on your device - welcome llama 3.2.
\newblock URL: \url{https://huggingface.co/blog/llama32}.

\bibitem[MS23]{maddigan2023chat2vis}
\textsc{Maddigan P., Susnjak T.}:
\newblock Chat2vis: generating data visualizations via natural language using chatgpt, codex and gpt-3 large language models.
\newblock \emph{Ieee Access 11} (2023), 45181--45193.

\bibitem[NJT24]{nagar2024zero}
\textsc{Nagar A., Jaiswal S., Tan C.}:
\newblock Zero-shot visual reasoning by vision-language models: Benchmarking and analysis.
\newblock In \emph{2024 International Joint Conference on Neural Networks (IJCNN)} (2024), IEEE, pp.~1--8.

\bibitem[Ope24]{openai2024gpt4technicalreport}
\textsc{OpenAI}:
\newblock Gpt-4 technical report, 2024.
\newblock URL: \url{https://arxiv.org/abs/2303.08774}, \href {http://arxiv.org/abs/2303.08774} {\path{arXiv:2303.08774}}.

\bibitem[OWJ{\etalchar{*}}22]{ouyang2022training}
\textsc{Ouyang L., Wu J., Jiang X., Almeida D., Wainwright C., Mishkin P., Zhang C., Agarwal S., Slama K., Ray A., et~al.}:
\newblock Training language models to follow instructions with human feedback.
\newblock \emph{Advances in neural information processing systems 35} (2022), 27730--27744.

\bibitem[PO23]{pandey2023mini}
\textsc{Pandey S., Ottley A.}:
\newblock Mini-vlat: A short and effective measure of visualization literacy.
\newblock In \emph{Computer Graphics Forum} (2023), vol.~42, Wiley Online Library, pp.~1--11.

\bibitem[PRS{\etalchar{*}}15]{pandey2015deceptive}
\textsc{Pandey A.~V., Rall K., Satterthwaite M.~L., Nov O., Bertini E.}:
\newblock How deceptive are deceptive visualizations? an empirical analysis of common distortion techniques.
\newblock In \emph{Proceedings of the 33rd annual acm conference on human factors in computing systems} (2015), pp.~1469--1478.

\bibitem[RKH{\etalchar{*}}21]{radford2021learning}
\textsc{Radford A., Kim J.~W., Hallacy C., Ramesh A., Goh G., Agarwal S., Sastry G., Askell A., Mishkin P., Clark J., et~al.}:
\newblock Learning transferable visual models from natural language supervision.
\newblock In \emph{International conference on machine learning} (2021), PMLR, pp.~8748--8763.

\bibitem[RNS{\etalchar{*}}18]{radford2018improving}
\textsc{Radford A., Narasimhan K., Salimans T., Sutskever I., et~al.}:
\newblock Improving language understanding by generative pre-training.

\bibitem[SEN24]{strobel2024hey}
\textsc{Str{\"o}bel M., Eckert K., Nagel T.}:
\newblock Hey chatgpt, can you visualize my data?--a multi-dimensional study on using an llm for constructing data visualizations.

\bibitem[SS23]{sultanum2023datatales}
\textsc{Sultanum N., Srinivasan A.}:
\newblock Datatales: Investigating the use of large language models for authoring data-driven articles.
\newblock In \emph{2023 IEEE Visualization and Visual Analytics (VIS)} (2023), IEEE, pp.~231--235.

\bibitem[SS24]{shah2024comprehensive}
\textsc{Shah M., Sureja N.}:
\newblock A comprehensive review of bias in deep learning models: Methods, impacts, and future directions.
\newblock \emph{Archives of Computational Methods in Engineering} (2024), 1--13.

\bibitem[TCD{\etalchar{*}}24]{tian2024chartgpt}
\textsc{Tian Y., Cui W., Deng D., Yi X., Yang Y., Zhang H., Wu Y.}:
\newblock Chartgpt: Leveraging llms to generate charts from abstract natural language.
\newblock \emph{IEEE Transactions on Visualization and Computer Graphics} (2024).

\bibitem[TLI{\etalchar{*}}23]{touvron2023llama}
\textsc{Touvron H., Lavril T., Izacard G., Martinet X., Lachaux M.-A., Lacroix T., Rozi{\`e}re B., Goyal N., Hambro E., Azhar F., et~al.}:
\newblock Llama: Open and efficient foundation language models.
\newblock \emph{arXiv preprint arXiv:2302.13971} (2023).

\bibitem[Vas17]{vaswani2017attention}
\textsc{Vaswani A.}:
\newblock Attention is all you need.
\newblock \emph{Advances in Neural Information Processing Systems} (2017).

\bibitem[V{\'a}z24]{vazquez2024llms}
\textsc{V{\'a}zquez P.-P.}:
\newblock Are llms ready for visualization?
\newblock In \emph{2024 IEEE 17th Pacific Visualization Conference (PacificVis)} (2024), IEEE, pp.~343--352.

\bibitem[WDX{\etalchar{*}}22]{wang2022fairness}
\textsc{Wang Z., Dong X., Xue H., Zhang Z., Chiu W., Wei T., Ren K.}:
\newblock Fairness-aware adversarial perturbation towards bias mitigation for deployed deep models.
\newblock In \emph{Proceedings of the IEEE/CVF conference on computer vision and pattern recognition} (2022), pp.~10379--10388.

\bibitem[WHB{\etalchar{*}}24]{wang2024aligned}
\textsc{Wang H.~W., Hoffswell J., Bursztyn V.~S., Bearfield C.~X., et~al.}:
\newblock How aligned are human chart takeaways and llm predictions? a case study on bar charts with varying layouts.
\newblock \emph{IEEE Transactions on Visualization and Computer Graphics} (2024).

\bibitem[WKR{\etalchar{*}}24]{wang2024human}
\textsc{Wang X., Kim H., Rahman S., Mitra K., Miao Z.}:
\newblock Human-llm collaborative annotation through effective verification of llm labels.
\newblock In \emph{Proceedings of the CHI Conference on Human Factors in Computing Systems} (2024), pp.~1--21.

\bibitem[WWS{\etalchar{*}}22]{wei2022chain}
\textsc{Wei J., Wang X., Schuurmans D., Bosma M., Xia F., Chi E., Le Q.~V., Zhou D., et~al.}:
\newblock Chain-of-thought prompting elicits reasoning in large language models.
\newblock \emph{Advances in neural information processing systems 35} (2022), 24824--24837.

\bibitem[ZGY{\etalchar{*}}24]{zou2024dynamath}
\textsc{Zou C., Guo X., Yang R., Zhang J., Hu B., Zhang H.}:
\newblock Dynamath: A dynamic visual benchmark for evaluating mathematical reasoning robustness of vision language models.
\newblock \emph{arXiv preprint arXiv:2411.00836} (2024).

\end{thebibliography}

\end{document}